# Ontology Based Feature Driven Development Life Cycle

**Farheen Siddiqui[1], M. Afshar Alam[1]**

**[1] Department of Computer Science, Hamdard University, New Delhi -110025 India**
`fsiddiqui@jamiahamdard.ac.in`
`aalam@jamiahamdard.ac.in`

### Abstract

The upcoming technology support for semantic web promises fresh directions for Software Engineering community. Also semantic web has its roots in knowledge engineering that provoke software engineers to look for application of ontology applications throughout the Software Engineering lifecycle. The internal components of a semantic web are "light weight", and may be of less quality standards than the externally visible modules. In fact the internal components are generated from external (ontological) component. That's the reason agile development approaches such as feature driven development are suitable for application's internal component development. As yet there is no particular procedure that describes the role of ontology in FDD processes. Therefore we propose an ontology based feature driven development for semantic web application that can be used form application model development to feature design and implementation. Features are precisely defined in the OWL-based domain model. Transition from OWL based domain model to feature list is directly defined in transformation rules. On the other hand the ontology based overall model can be easily validated through automated tools. Advantages of ontology-based feature Driven development are also discussed.

## 1. Introduction

The Software Engineering and Knowledge Engineering groups work on overlapping domain. Software Engineering people pay more attention to software modeling and Knowledge Engineering community has come up with variety of modeling approaches in order to realize the vision of the semantic web [1].Semantic web has made this overlap even more wide but still there is less forums for discussing synergies is (e.g. SWESE1, SEKE2 and W3C3) .The methods on integrating Software and Knowledge Engineering approaches focus on approaches of meta-modeling, but are abstract for software engineers in terms of there application in software process. Current approaches of modeling only partially solve the problem related to component reuse, composition, validation, information and application integration, software testing and quality. Such basic needs are generating new approaches towards every single aspect in software engineering.

Domain analysis is an essential activity for successful reuse across applications in the same domain. Domain model is essential for domain and application-specific development. And therefore should meet some requirements. First, it should provide guidance for the design of architecture and components. Second, the model should provide means to get validated against system constraints. Third, it should be customizable for specific application. In "semantic web" era, developer would discover shareable domain models and knowledge bases from a variety of interrelated repositories and then connect them together with application specific components. Thus all applications that share overlapping domain models would then have a certain degree of interoperability built in. These sharable domain models are referred as domain ontology and provide many benefits such as model reuse, flexibility, consistency checking and reasoning. Also new technologies and tools have been developed for ontology representation, machine-processing, and ontology sharing. This makes their adoption in real-world applications much easier. While ontologies are about to enter mainstream Software Engineering practices, their applications in software engineering are manifold. Despite of using a well defined domain model it is not uncommon for software projects to exceed budget, blow schedule, and deliver something less than desired .The main reason behind this is the scenario of ever changing user requirement and lack of communication between customer and developer team. Therefore a process for delivering frequent, tangible, working results is most desired. Agile development approaches focus on these issues and feature driven development is among one of the approaches towards it. The remainder of this paper is organized as follows. Section 2 presents the ontology based feature driven development and the stages involved in it. Section 3 introduces the ontology-based "overall model", Section 4 elaborates the process of feature list development and planning and section 5 discuss about component development. Finally, we draw our conclusions with discussion of ontology-based feature modeling and future work in section 6.

## 2. Feature Driven Development

Feature Driven Development is a model-driven short-iteration process. It begins with establishing an "overall model" shape. Then it continues with a series of two week "design by feature, build by feature" iterations. The features are small "useful in the eyes of the client" results. Iteration like "build the admission subsystem" would take too long to complete. Iteration like "build the data access layer" is not exactly client-valued. In contrast, a small feature like "assign unique enrollment number" is both short and client-valued. FDD is based on its first process of developing overall model. This process is so critical that it is referred as process 1 in FDD life cycle. Therefore a strictly defined modeling basis for "overall model" is essential, which should provide a mechanism to connect model elements in various development phases. Ontology related theory is a suitable way to achieve our goals. Ontology is a conceptualization of a domain or subject area typically captured in an abstract model of how people think about things in the domain [2]. Rubén [2] considers domain models as narrow or specialized ontology, and the main difference is that domain models define abstract concepts in an informal way and have no axioms. Because of the facilities for the generalization and specialization of concepts and the unambiguous terminology it provides [3], ontology has been widely used in domain knowledge representation and requirement modeling, reuse and consistency checking. For example, Sugumaran etc. [4] proposed a semantic-based approach to component retrieval, in which ontology and domain models are adopted for capturing application domain specific knowledge to express more pertinent queries for component retrieval. Girardi etc. [3] proposed GRAMO, an ontology based technique for the specification of domain and user models in Multi-Agent domain engineering.

The purpose of this paper is to reduce the gap between knowledge engineering and software engineering by using ontology in every step of a FDD process.This paper proposes an ontology-based feature driven development methodology, in which OWL ontology is considered as an overall model and is used at every step of FDD. In this way, we can provide better support for domain modeling, and succeeding domain design and implementation. First, ontology-based feature model can be formally represented easily and validation of the model can be realized through ontology reasoning. Second, the ontology-based unambiguous terminology provide precise and detailed semantic knowledge for the domain, so the ontology based feature model can also be adopted as the domain business model and contain enough information for component description and architecture design.

## 3. Ontology Based Feature Driven Development

In this section, we will present method of using ontologies in the context of FDD. The presentation will be in the order of FDD life cycle as described in fig 1. In each step we will discuss how ontology can be used and what benefits we can achieve by its usage.

Traditionally FDD life cycle is based on following five processes.:

Process #1: Develop an overall model (using initial requirements/ features, snap together with components, focusing on shape).

Process #2: Build a detailed, prioritized features list.

Process #3: Plan by feature.

Process #4: Design by feature (using components, focusing on sequences).

Process #5: Build by feature.

For ontology based feature driven development we have merged these into three stages as depicted in Fig 1. At each stage ontology is used as the basic building block. Software modeling languages and methodologies can benefit from the integration with ontology languages such as RDF and OWL in various ways, e.g. by reducing language ambiguity, enabling validation and automated consistency checking Ontology languages provide better support for logical inference, integration and interoperability than MOF-based languages. UML-based tools can be extended more easily to support the creation of domain vocabularies and ontologies. Since ontologies promote the notion of identity, Ontology Definition Metamodel and related approaches simplify the sharing and mediation of domain models. Since a domain model is initially unknown and changes over time, a single abstraction and separation of concerns is considered feasible if not necessary Therefore a single representation of the domain model should be shared by all participants throughout the lifecycle to increase quality and reduce costs. The mapping of a domain model to code should therefore be automatized to enable the dynamic use by other components and applications. Fig 1 depicts the three dimensional view of FDD life cycle that uses semantic web technologies at each stage of development. FDD begins with application model development. We use OWL and SWRL to define the entities, classes, hierarchies and domain rules in form of problem ontology. At second process the feature list is generated from the problem ontology and planning is done with SQWRL[15].At final process of component building each feature generated from problem ontology is designed and implemented using APIs like jena.In the following

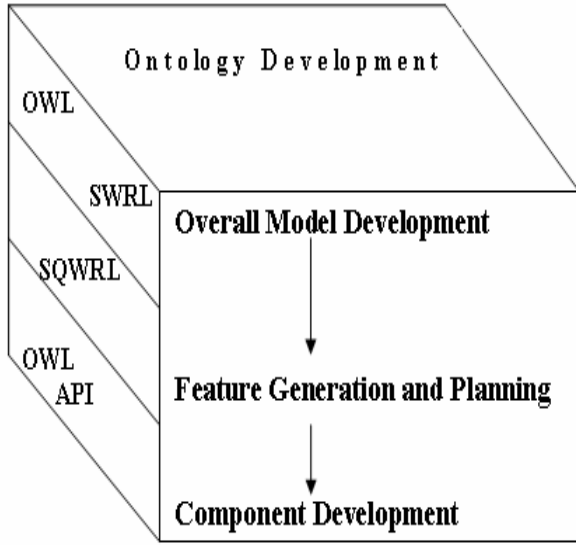

Fig. 1  FDD Life Cycle Model.

## 4. Develop an Overall Model

Developing the initial model shape needs involvement of both domain and development members. Domain member starts with presenting an abstract view and scope of the system within application context. The domain and development members develop a rough model that can be followed at the initial stage. Later on the domain and development member stepwise explores each detail aspect of the system and merge the understanding in the initial model alongside adjusting model shape. The development of overall model starts when the client is ready to proceed with the building of a system but he might not express the requirement in any concrete format. Hence at first this phase deals with gathering the desired system functionality from the customers. Since the involved software engineers are often no domain experts, they must learn about the problem domain from the customers. A different understanding of the concepts involved may lead to an ambiguous, incomplete specification and major rework after system implementation. Therefore it is important to assure that all participants in the phase have a shared understanding of the problem domain. Moreover, change of requirements needs to be considered because of changing customer's objectives.

An ontology can be used for both, to describe requirements specification documents [5, 6] and formally represent requirements knowledge [7,8]. Ontologies can cover semi-formal and structured as well as formal representation [7]. Further, the "domain model" represents the understanding of the domain under consideration, i.e. in the form of concepts, their relations and business rules. It is formalized using a conceptual modeling language such as the UML. Moreover, the problem domain can be described using an ontology language, with varying degrees formalization and expressiveness. In contrast to traditional knowledge-based approaches, e.g. formal specification languages, ontologies seem to be well suited for an evolutionary approach to the specification of requirements and domain knowledge [7] that is needed to achieve agility in development cycle. Moreover, ontologies can be used to support requirements management and traceability [6]. Automated validation and consistency checking are considered as a potential benefit compared to semi-formal or informal approaches providing no logical formalism or model theory. Finally, formal specification may be a prerequisite to realize model-driven approaches in the design and implementation phase. At the end of the process 1, an overall ontology based model is developed and based on that an informal feature list is noted down. In this paper to support the life cycle, we have taken an example of University system. Following the above procedure the developer and domain expert build Education ontology in Protégé .Fig 2 shows graphical representation of Education ontology developed in Protégé.

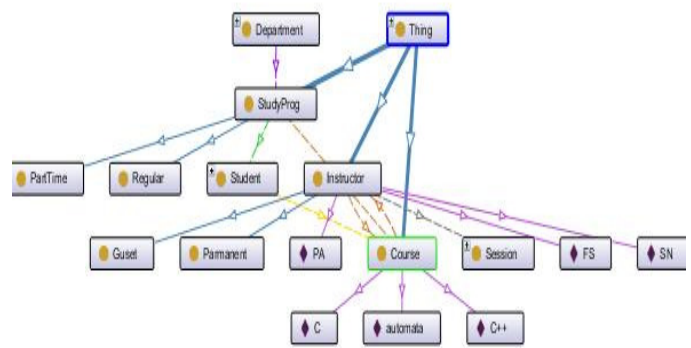

Fig. 1  Education Ontology in Protégé.

Ontologies are purely conceptual models that capture domain concepts and neglects domain-restricted rules. If the requirements model violate these rules or contradict the usual business behavior, they become unreasonable. We have used SWRL to model the integrity rules and derivation rules which restrict the business behavior. The requirements model represented by domain ontology can be checked for consistency using HermiT reasoners and rules can be checked with Jess rule Engine.Thus model development process is both guided by domain ontology and restricted by domain rules. Therefore, the model would comply with both business needs and domain knowledge. A rule contains one or more antecedent and one consequent, the description is as follows:

```
<owlr : rule rdf : ID ="rule ID">
<owlr : annotation >ruleName</ owlr :
annotation >
<owlr :body>antecedent</ owlr :body>
<owlr :head>consequent</ owlr :head>
</ owlr : rule>
```

Consider the rule that a person having a salary associated is an employee. This rule can be ststed in SWRL as:

```
Person(?p)    ^    salary(?p,   ?s)   ->
Employee(?p)
```

and encoded in ontology as

```
DLSafeRule>
<Body>
<ClassAtom>
<Class IRI="#Person"/>
<Variable IRI="urn:swrl#p"/>
</ClassAtom>
<DataPropertyAtom>
<DataProperty IRI="#salary"/>
<Variable IRI="urn:swrl#p"/>
<Variable IRI="urn:swrl#s"/>
</DataPropertyAtom>
</Body>
<Head>
<ClassAtom>
<Class IRI="#Employee"/>
<Variable IRI="urn:swrl#p"/>
</ClassAtom>
</Head>
</DLSafeRule>
```

# 6. Feature Generation and Planning

While building the feature list, the main task is to identify the features, groups them hierarchically, prioritizes them, and weights them. In subsequent iterations of this process, smaller teams tackle specialized feature areas. We propose to establish one to one correspondence between the ontology and the feature list development. We can use the ontology develop at previous step to generate features supported by it and can also group features into feature set.

## 5.1 Feature List Generation

The process starts with the informal features list from FDD Process 1. It then:
_ transforms object property in the ontology into features of their domain,
_ transforms classes in the ontology into feature sets
We can use the formats:
_ For features: <action> the <object property-range> <by|for|of|to> a(n) <Class-name>
_ For feature sets: <Class-name> module including all subclass of <Class-name>
_ For major feature sets: <ontology-name> management
For example in Education Ontology classes can be Student , Department ,StudyProgram, Courses, Session ,Attendance ,Instructor etc.Also a object property hasStudyProg has domain of Department and range of StudyProg. Therefore a feature: department offers study program can be considered as a feature in form
Offering of StudyProgram by Department, which is a triple of form:
<action> <object property range> by|for|of|to <object property domain>
This can be inferred from ontology as hasStduyProg is a object property of Department and this feature belongs to department module of education management. To exit this process, the features-list team must deliver a detailed features list, grouped into major feature sets and feature sets.

## 5.2 Feature Planning

At planning stage the project manager, the development manager, and the chief programmers establish milestones The planning team determines the development sequence and sets initial completion dates for each feature set and major feature sets for "design by feature, build by feature" iterations. Using the development sequence and the feature weights as a guide, the planning team assigns classes to class owners. Using the development sequence and the feature weights as a guide, the planning team assigns chief programmers as owners of feature sets (classes in ontology). Every class in ontology can be associated with a property of "hasowner".A feature indicates the class(es) involved and a query can be framed in SQWRL to fetch the class owner of corresponding classes in ontology. For example to find out owner of a particular class Instructor for feature "assign Course to Instructor" the following query can be  used:
Course(?c) ^ Instructor(?I) ^ hasCourse(?I, ?c) ^ hasOwner(I,P)-> sqwrl:select(?I, ?P)

To exit this process, the planning team must produce a development plan, subject to review and approval by the development manager and the chief architect. The plan consist of an overall completion date, for each major feature set, and feature: its owner and its completion date , for each class, its owner.

# 4. Component Development

This stage consists of iterations feature design, feature implementation.

## 5.2 Feature Design

A chief programmer takes the next feature, identifies the classes likely to be involved, and contacts the corresponding class owners. This feature team works out a detailed sequence diagram. Chief programmer identifies the classes likely to be involved in the design of this feature and identifies the developers needed to form the feature team. He contacts those class owners, initiating the design of this feature.While developing the design the team also can look for components that already exist when implementing functionality, since reuse can avoid rework, save money and improve the overall system quality. Usually, this search for reusable components takes place after the analysis phase, when the functional requirements are settled [9].  Ontologies can help here to describe the functionality of components using a knowledge representation formalism that allows more convenient and powerful querying [10]. One approach implementing this is the KOntoR system that allows storing semantic descriptions of components in a knowledge base and running semantic queries on it. Compared to traditional approaches, ontologies provide two advantages in this scenario. First, they help to join information that normally resides isolated in several separate component descriptions. Second, it provides background knowledge that allows non-experts to query from their point of view .

## 5.2 Feature Implementation

Each class owner builds his object property for the feature. He extends his class-based test cases and performs class-level (unit) testing. Once the code is successfully implemented and inspected, the class owner checks in his class(es) to the configuration management system. When all classes for this feature are checked in, the chief programmer promotes the code to the build process.

At the end of this phase, the feature team must delivers implemented and inspected classes and properties with unit testing. The mapping of a domain model to code should be automated to enable the dynamic use by other components and applications. The programmatic access of ontologies and manipulation of knowledge bases using ontology APIs requires special knowledge by the developers. Therefore an intuitive approach for object-oriented developers is desirable [cf. 23]. This can be achieved by ontology tools that generate an API from the ontology, e.g. by mapping concepts of the ontology to classes in an object oriented language. The generated domain object model can then be used managing models, inferencing, and querying. Tools supporting those features are already available today, e.g. [12] and [13].The domain model encoded in OWL can be used at implementation time with OWL API.

Semantic Web applications usually need to make some ontological commitments, i.e., they need to have hard-coded knowledge about a certain domain ontology. In the example above, the application has hard-coded behavior that depends on the education.owl ontology, which contains classes like **Instructor** and **Course**. The application can exploit reasoning engines like Racer or rule engines like SWRL to expose "intelligent" behavior. All of this is controlled by some logic (in this example it is Java code), which also interacts with the end user by means of interface technologies like JSPs, Swing applications, or Web Services. Protege-OWL API features can be used for developing stand-alone applications. Such applications can load ontologies from the Semantic Web, perform queries on them, add or edit resources from the ontology, classify instances and classes, and write out resulting ontologies to a file. From an object-oriented perspective, Owl API can generate code for class such as:

```
public interface Person {
    String getFirstName();

    void setFirstName(String value);
    ...

}
```
so that we can use code like this:
```
public interface Person extends
OWLIndividual {

    String getFirstName();
    void setFirstName(String value);
    ...
}
```

and then provide a default implementation like the following scheme:

```
public class DefaultPerson extends
DefaultOWLIndividual implements Person
{

    public DefaultPerson(KnowledgeBase
kb, FrameID id) {
        super(kb, id);
    }
```

```
    public String getFirstName() {
        RDFProperty property =
getOWLModel().getRDFProperty("firstName
");
        return (String)
getPropertyValue(property);
    }

    public void setFirstName(String
value) {
        RDFProperty property =
getOWLModel().getRDFProperty("firstName
");
        setPropertyValue(property,
value);
    }

    ...
}
```

## 7. Conclusions

A strictly-defined formal basis is essential for applicable domain modeling. In this paper, ontology is used as the foundation of the FDD life cycle. Ontology has been widely adopted in domain knowledge modeling and has corresponding modeling language, such as OWL.Furthermore, rule-based reasoning can be performed on the ontology model for model validating. Establishing a mapping between domain model and the architecture is the objective of domain engineering [14]. However, there is a large gap between the domain model representation and actual implementation. We can reduce the gap by establishing a smooth transition from elements in the domain model (i.e. features) to elements in the architecture model (i.e. components). In our approach, domain ontology (i.e. the ontology-based overall model) is also representation basis for component semantics. Our future work will be based on the complete implementation of an education system through feature driven development using education ontology. Also in future we will develop an ontology based architecture and design pattern for semantic web application.